\documentclass[twocolumn]{jpsj2}

\title{Charge-Transfer Excitations in One-Dimensional Dimerized Mott Insulators}
\author{
Nobuya \textsc{Maeshima}${}^{1,2}$\thanks{E-mail address: maeshima@ims.ac.jp} and
Kenji \textsc{Yonemitsu}${}^{2,3}$
}
\inst{
${}^1$ Department of Chemistry, Tohoku University, Aramaki, Aoba-ku, Sendai 980-8578\\
${}^2$ Institute for Molecular Science, Okazaki 444-8585 \\
${}^3$ Department of Functional Molecular Science, Graduate University for Advanced Studies, Okazaki 444-8585 
}

\recdate{\today}

\abst{
We investigate the optical properties of one-dimensional (1D) dimerized Mott insulators using the 1D dimerized extended Hubbard model.  Numerical calculations and a perturbative analysis from the decoupled-dimer limit clarify that there are three relevant classes of charge-transfer (CT) states generated by photoexcitation: interdimer CT unbound states, interdimer CT exciton states, and intradimer CT exciton states.   This classification is applied to understanding the optical properties of an organic molecular material, 1,3,5-trithia-2,4,6-triazapentalenyl (TTTA), which is known for its photoinduced transition from the dimerized spin-singlet phase to the regular paramagnetic phase.  We conclude that the lowest photoexcited state of TTTA is the interdimer CT exciton state and the second lowest state is the intradimer CT exciton state.
}

\kword{strongly correlated electron systems, spin-Peierls transition, photoinduced phase transition, optical conductivity}
\begin{document}
\maketitle

\section{Introduction}

Currently, there is extensive interest in photoinduced phenomena of strongly correlated electron systems.~\cite{yu,PrMnO3,VO2,TaS2,TTFCA1,TTFCA2,poly,MX,MMX,EDO,sitaET,KTCNQ1,KTCNQ2,TTTA1,takeda}  These phenomena show drastic changes in magnetic or electronic properties triggered by photoirradiation, and often accompany subsequent ultrafast relaxation to equilibrium states.  To reveal the fundamental physics of these phenomena is one of important challenges in condensed matter physics, and furthermore it would be of great help for the application of related materials to new optical switch devices.

Theoretical efforts have also been devoted to this field recently.~\cite{nasu,yonemitsu1,hanamura,nagaosa1,koshino1,huai,yonemitsu2,iwano,yonemitsu3,maeshima1,maeshima3}  From the theoretical point of view, the chief ingredient to understand the photoinduced phenomena is to clarify the nature of excited states relevant to the photoexcitation.  Photoirradiation generates optically allowed excited states (photoexcited states), introducing novel characteristics to the system.  Some photoexcited state may lead the system to an unconventional macroscopic phase that is quite different from any thermodynamic one.  Thus, to identify the relevant photoexcited states is the starting point of the theoretical investigation for how the photoinduced phenomena occur.

In this paper, we present a simple picture to describe the photoexcited states of a particular class of strongly correlated electron systems, one-dimensional (1D) dimerized Mott insulators.  This picture is derived from a perturbative treatment from the decoupling limit of the 1D dimerized extended Hubbard model and its numerical calculations.  We demonstrate that there are three types of photoexcited charge-transfer (CT) states: (i) interdimer CT unbound states, (ii) interdimer CT exciton states, and (iii) intradimer CT exciton states.  This theoretical understanding is applied to discussion about the experimental results of  an organic radical crystal, 1,3,5-trithia-2,4,6-triazapentalenyl (TTTA).~\cite{TTTA2}
This material is a quasi-1D dimerized Mott insulator below the spin-Peierls transition temperature, and shows a characteristic photoinduced transition from the dimerized spin-singlet phase to the regular paramagnetic phase.~\cite{TTTA1,takeda}
In addition, its optical excitations have not completely been understood.~\cite{TTTA1,fujita} In this work, we demonstrate that the lowest photoexcited state of TTTA is the interdimer CT exciton state and the intradimer CT exciton has a higher excitation energy.

\section{Model}

In this work, we use the 1D dimerized extended Hubbard model for 1D dimerized Mott insulators.  The Hamiltonian is given by
\begin{eqnarray}
{\cal H} &=& -t\sum_{\sigma}\sum_{l=0}^{N-1}[1+(-1)^l\delta]
( c^\dagger_{l+1,\sigma}c^{}_{l,\sigma} +  c^\dagger_{l,\sigma}c_{l+1,\sigma} )
\nonumber \\ &+& U\sum_l n_{l,\uparrow}n_{l,\downarrow} 
+ V\sum_{l} n_{l}n_{l+1},
 \label{eq:ham}
\end{eqnarray}
where $c^{\dagger}_{l,\sigma}$ ($c_{l,\sigma}$) is the creation (annihilation) operator of an electron with spin $\sigma$ at site $l$, and $n_{l,\sigma}=c^{\dagger}_{l,\sigma}c_{l,\sigma}$.  The nearest-neighbor transfer integrals alternate as $t(1+\delta),t(1-\delta),\cdots$, and the parameter $t$, the transfer integral of the regular lattice, is used as a unit of energy in this paper.  The band-filling of electrons is set to be a half.
The Coulomb repulsion is taken into account up to the nearest neighbor: the on-site Coulomb interaction $U$ and the nearest-neighbor interaction $V$. 

\section{Analysis from the Decoupling Limit}

In this section, we invoke the analysis from the decoupled-dimer limit of the model~(\ref{eq:ham}).~\cite{soos,mukho,maeshima3} This analysis is able to treat $U/t$ exactly, and is proved to be appropriate to construct a quite simple picture of the photoexcited states.  Although several studies have also presented different approximations to discuss the optical properties of this model,~\cite{lyo,gallagher,gebhard1,gebhard2} these preceding theories are justified only in the $U/t\to\infty$ limit.  In fact, our treatment also has a disadvantage; the validity of the approximation is not confirmed for small $\delta$.
However, in this work, we can restrict ourselves to systems with rather large dimerization, typically with $\delta(\le 0.4)$, because the organic material TTTA is considered to have a quite large value of $\delta \sim 0.46$.~\cite{fujita}

In our preceding paper,~\cite{maeshima3} this analysis for $V=0$ has provided a clear classification of the photoexcited charge-transfer (CT) states: intradimer CT states and interdimer CT states.   In the following, we discuss how this classification is influenced by finite $V$.  All the notations used here are defined in ref.~\citen{maeshima3}, with a replacement from $U$ to $U'\equiv U-V$.

\subsection{Model}

Following the formulation in ref.~\citen{maeshima3}, we split the Hamiltonian~(\ref{eq:ham}) into the intradimer part ${\cal H}_0$ and the interdimer part ${\cal H}_1$ as follows:
\begin{equation}
{\cal H}= {\cal H}_0 + {\cal H}_1,
\end{equation}
with
\begin{eqnarray}
{\cal H}_0 &=& -t(1+\delta)\sum_{\sigma}\sum_{l\in \rm e}
( c^\dagger_{l+1,\sigma}c_{l,\sigma} +  c^\dagger_{l,\sigma}c_{l+1,\sigma} \nonumber )\\
 &&+  U\sum_{\sigma}\sum_{l\in \rm e} (n_{l,\uparrow}n_{l,\downarrow} +  n_{l+1,\uparrow}n_{l+1,\downarrow})\nonumber\\
 &&+  V\sum_{l\in \rm e} n_ln_{l+1}
\end{eqnarray}
and
\begin{eqnarray}
{\cal H}_1 &=& -t(1-\delta)\sum_{\sigma}\sum_{l\in \rm o}
( c^\dagger_{l+1,\sigma}c_{l,\sigma} +  c^\dagger_{l,\sigma}c_{l+1,\sigma}  ) \nonumber \\
&&+V\sum_{l\in \rm o} n_ln_{l+1},
\end{eqnarray}
where $\sum_{l \in \rm e(o)}$ denotes the summation over even (odd) $l$.  In what follows, ${\cal H}_0$ is considered to be the 0-th order Hamiltonian and ${\cal H}_1$ be the perturbation term.  Hence this treatment is valid for $(1-\delta),V/t<<1$.

Before discussing photoexcited CT states, we identify the ground state of the 0-th order system.  The ground state $|\psi_0^0\rangle$ of ${\cal H}_0$ is a direct product of the ground states of isolated dimers,
\begin{equation}
|\psi_0^0\rangle = |G\rangle_{0} \otimes |G\rangle_{1} \otimes \cdots \otimes |G\rangle_{N_d-1},
\end{equation}
where the state $|G\rangle_{n}$ is the ground state of the $n$-th dimer ($n=0,1,\cdots,N_d-1$) and 
$N_d(\equiv N/2)$ is the number of dimers.  We note that for a strongly correlated system ($U'>>t$), $|G\rangle_{n}$ is equal to the so-called spin-singlet dimer $(|\uparrow\downarrow\rangle-|\downarrow\uparrow\rangle)/\sqrt{2}$.

\subsection{Intradimer CT state and interdimer CT state}

The intradimer CT state is generated by an intradimer charge transfer.  By this charge transfer on the dimer $n$, the dimer ground state $|G\rangle_n$ is converted to $|O^2\rangle_n$ defined by
\begin{equation}
|O^2\rangle_n = \frac{1}{\sqrt{2}} ( c^\dagger_{2n \uparrow} c^\dagger_{2n \downarrow}
- c^\dagger_{2n+1 \uparrow} c^\dagger_{2n+1 \downarrow}  ) |0\rangle.
\end{equation}
 The corresponding excitation energy is given by
\begin{equation}
\Delta_{\rm intra} = U'/2 + \sqrt{U'^2/4 + 4t'^2 },
\label{eq:gap_intra}
\end{equation}
where $t'=t(1+\delta)$.  We note that the state $|O^2\rangle_n$ has an exciton-like character because the holon and the doublon~\cite{mizuno} are bound in a single dimer.

The interdimer CT state is generated by an interdimer charge transfer.
By this charge transfer between the dimers $n$ and $n+1$, the two dimer ground states $|G\rangle_n|G\rangle_{n+1}$ are converted to $|O^3_{\sigma}\rangle_n|E^1_{\bar{\sigma}}\rangle_{n+1}$ or $|E^1_{\sigma}\rangle_n|O^3_{\bar{\sigma}}\rangle_{n+1}$.
 Here, $|O^3_{\sigma}\rangle_n$ and $|E^1_{\bar{\sigma}}\rangle_{n}$ are eigenstates of the isolated dimer $n$ and defined by
\begin{eqnarray}
|O^3_{\sigma}\rangle_n &=& \frac{1}{\sqrt{2}} ( c^\dagger_{2n\bar{\sigma}} c^\dagger_{2n+1\sigma}  c^\dagger_{2n\sigma} 
+ c^\dagger_{2n+1\bar{\sigma}}  c^\dagger_{2n+1\sigma}  c^\dagger_{2n\sigma}  ) |0\rangle, \nonumber \\
|E^1_{\bar{\sigma}}\rangle_n &=& \frac{1}{\sqrt{2}} ( c^\dagger_{2n\bar{\sigma}} + c^\dagger_{2n+1\bar{\sigma}} )|0\rangle,
\end{eqnarray}
where $\sigma$ and $\bar{\sigma}$ are spin indices and $\bar{\sigma}$ is opposite to $\sigma$.  Its excitation energy is given by
\begin{equation}
\Delta_{\rm inter} = V - 2t' + 2\sqrt{U'^2/4 + 4t'^2 }.
\label{eq:gap_inter}
\end{equation}
In contrast to the intradimer CT state, the interdimer CT state has a free-carrier-like character for $V/t<<1$. For $V=0$, $|O^3_{\sigma}\rangle$ and $|E^1_{\bar{\sigma}}\rangle$ move freely by the 1st-order process of ${\cal H}_1$.~\cite{maeshima3}

For $V/t<<1$, the free-carrier-like interdimer CT state is the lowest photoexcited state and the exciton-like intradimer CT state has a higher excitation energy.~\cite{soos,mukho,maeshima3}  For $U'>>t'$, the two excitation energies are given by
\begin{equation}
 \Delta_{\rm intra} \sim U-V,
\label{eq:delta_intra}
\end{equation}
and
\begin{equation}
 \Delta_{\rm inter} \sim U-2t(1+\delta).
\label{eq:delta_inter}
\end{equation}
Thus, the reduction in the energy of the interdimer CT state is caused by the kinetic energy gain due to the holon in $|E^1_{\sigma}\rangle$ and the doublon in $|O^3_{\bar{\sigma}}\rangle$.

\subsection{Bound-unbound transition of the interdimer CT state}

\begin{figure}[hbt]
\begin{center}
\includegraphics[width=5.5cm,clip]{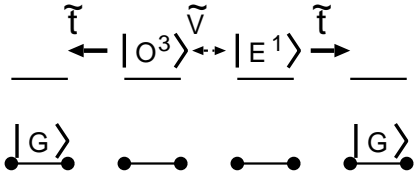}
\end{center}
\caption{Schematic picture of ``particles'' in the interdimer CT state.}
\label{fig:opt1}
\end{figure}

Here we demonstrate that small but finite $V$ causes a bound-unbound transition of the interdimer CT state.
As shown in the previous subsection, the lowest photoexcited state for $V/t<<1$ is the interdimer CT unbound state that consists of free ``particles'' $|O^3_{\sigma}\rangle$ and $|E^1_{\bar{\sigma}}\rangle$.
Thus, it is reasonable that, for small but finite $V/t$,  $|O^3_{\sigma}\rangle$ and $|E^1_{\bar{\sigma}}\rangle$ are bound, resulting in an interdimer CT exciton state.
To discuss this point, we here use the following basis set:~\cite{maeshima3}
\begin{eqnarray}
|n_1,n_2\rangle &\equiv& \frac{1}{2}
\left[ |O^3_{\uparrow} E^1_{\downarrow}\rangle_{n_1,n_2} + 
|O^3_{\downarrow} E^1_{\uparrow}\rangle_{n_1,n_2} \right. \nonumber \\
 && \left. - |E^1_{\uparrow} O^3_{\downarrow}\rangle_{n_1,n_2}
- |E^1_{\downarrow} O^3_{\uparrow}\rangle_{n_1,n_2} \right].
\end{eqnarray}
Here the state $|XY\rangle_{n_1,n_2}$, with $(X,Y)=(O^3_{\sigma}, E^1_{\bar{\sigma}})$ or $(E^1_{\sigma}, O^3_{\bar{\sigma}})$,  means that $|X\rangle$ is located on the $n_1$-th dimer, $|Y\rangle$ lies on the $n_2$-th ($n_1<n_2$) dimer, and the other dimers are in the singlet-dimer state $|G\rangle$. 
  Without ${\cal H}^1$, $|n_1,n_2\rangle$ are degenerate for arbitrary pairs of $(n_1<n_2)$.  Then we carry out the degenerate perturbation theory to treat the first-order effect of ${\cal H}^1$, and obtain
\begin{equation}
{\cal H}_1|n_1,n_2\rangle = \tilde{t} \sum_{\tau=-1,1}
( |n_1+\tau,n_2\rangle + |n_1,n_2+\tau\rangle)
\end{equation}
for $n_2-n_1\ge 2$, and 
\begin{equation}
{\cal H}_1|n_1,n_2\rangle = \tilde{t}
(  |n_1-1,n_2\rangle + |n_1,n_2+1\rangle ) - \tilde{V} |n_1,n_2\rangle 
\end{equation}
 for $n_2-n_1 = 1$, where
\begin{eqnarray}
\tilde{t}&=&t(1-\delta)(\alpha+\beta)^2/4 \quad {\rm ,} \alpha = 2t'/C, \quad, \nonumber \\
 \beta &=& ( \sqrt{U'^2/4+4t^2 } -   U'/2 )/C, \nonumber \\
 \quad C&=& (U'^2/2 + 8t'^2 -U'\sqrt{U'^2/4+4t'^2 }           )^{1/2},
\end{eqnarray}
and
\begin{equation}
\tilde{V}=V/4.
\end{equation}
This Hamiltonian is equivalent to that of a two-spinless-fermion system defined by
\begin{equation}
 \tilde{\cal H} = \sum_l [ \tilde{t}(a^\dagger_l a_{l+1} + a^\dagger_{l+1} a_{l})  - \tilde{V} n_l n_{l+1} ],
\label{eq:ff}
\end{equation}
and the ``particles'' move with the transfer integral $\tilde{t}$ and attract with each other via $\tilde{V}$ (see Fig.~\ref{fig:opt1}).
The attractive interaction $\tilde{V}$ is caused by the attractive force between the holon in $|E^1\rangle$ and the doublon in $|O^3\rangle$.
The Bethe ansatz~\cite{bethe} of eq.~(\ref{eq:ff}) tells us that the bound-unbound transition occurs at $\tilde{V}=2\tilde{t}$, resulting in
\begin{equation}
V_{\rm b}/t=2(1-\delta)(\alpha+\beta)^2.\label{eq_Vb}
\end{equation}
Therefore for $V \le V_{\rm b} (>V_{\rm b})$ the lowest state is an unbound state (a bound state), and the bound state is the interdimer CT exciton state. 

\subsection{Summary of the photoexcited CT states}

Our discussion has demonstrated that there are three types of photoexcited CT states:
\begin{enumerate}
\item The interdimer CT unbound state: when $V<V_b$, $|O^3\rangle$ and $|E^1\rangle$, generated by the interdimer charge-transfer, move freely in the whole system.
\item The interdimer CT exciton state:  when $V$ larger than $V_b$ is introduced, $|O^3\rangle$ and $|E^1\rangle$ are bound.
\item The intradimer CT exciton state: the holon and the doublon are strongly bound in a single dimer.
\end{enumerate}
These three types of CT states have been mentioned in a model with longer-range repulsive interactions, associated with nonlinear optical responses of conjugated polymers.~\cite{soos}
Here we demonstrate that these three states are important for the systematic understanding of the photoexcited states in molecular crystals including TTTA, where models with short-range interactions are appropriate.~\cite{seo}
In addition, we show that their wave functions and relative energies are governed by $V/t$.

Although this classification is basically valid for small $1-\delta,V/t$ where the perturbative treatment of ${\cal H}_1$ is appropriate, 
for large $V$, a simple speculation can be obtained as below.  As $V$ increases, the intradimer CT exciton state linearly decreases its excitation energy.  On the contrary, the interdimer CT state is less sensitive to $V$.  As a result, a level crossing would occur between them and then the lowest photoexcited state would change from the interdimer CT state to the intradimer CT state.  In fact, there is a crossover of the lowest photoexcited state, from the interdimer CT state to the intradimer CT state because two CT states have the mixing term,~\cite{mukho}
\begin{equation}
_n\langle O^2| _{n+1}\langle G| {\cal H}_1|O^3\rangle_n |E^1\rangle_{n+1} = t(1-\delta)(\alpha+\beta).
  \label{eq:cross}
\end{equation}

\begin{figure}[hbt]
\begin{center}
\includegraphics[width=5.5cm,clip]{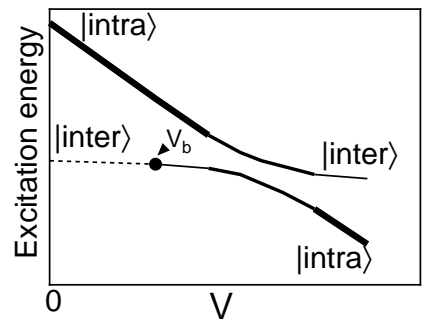}
\end{center}
\caption{Energy diagram of relevant excited states. The solid (dotted) lines show the exciton (unbound) states, and the widths of the lines indicate the strength of the spectral weight.}
\label{fig:sche}
\end{figure}

This argument is summarized in a schematic energy diagram of Fig.~\ref{fig:sche}.  When $V<V_b$, the lowest photoexcited state is the interdimer CT unbound state with $|O^3\rangle$ and $|E^1\rangle$ while the intradimer CT exciton has a higher excitation energy.  As for the spectral weight, as shown later, the intradimer CT exciton is dominant and the interdimer CT unbound state has a small contribution.
  As $V$ increases, the interdimer CT state becomes the exciton state with bound $|O^3\rangle$ and $|E^1\rangle$, although its spectral weight would remain small.  As $V$ further increases, the intradimer CT exciton lowers its energy level and approaches the interdimer CT exciton state, resulting in the crossover between them.  Then the two states have comparable spectral weights because of the mixing.  For much larger $V$, the lowest photoexcited state becomes the intradimer CT exciton state and has the dominant spectral weight.

\section{Numerical Results}

  \subsection{Optical conductivity}

In this section, we examine whether our analytical discussion is appropriate or not for the realistic parameter region.  To find photoexcited states of the model~(\ref{eq:ham}), we calculate the optical conductivity spectrum of the ground state $|\psi_{0}\rangle$ given by
\begin{equation}
\sigma(\omega)= \frac{\pi}{N\omega} \sum_i |\langle\psi_i|J|\psi_0\rangle|^2 \delta(\omega+E_{0}-E_i),
\label{eq:sigma-reg}
\end{equation}
where
\begin{equation}
J\equiv it\sum_{l,\sigma}[1+(-1)^l\delta]( c^\dagger_{l+1,\sigma}c_{l,\sigma} -  c^\dagger_{l,\sigma}c_{l+1,\sigma})
\end{equation}
is the current operator, $|\psi_i\rangle$ the $i$-th eigenstate, and $E_i$ the corresponding energy.  The lattice constant is set to be unity for simplicity.  We note that the Drude part is omitted because the model~(\ref{eq:ham}) is treated only in the insulating phase.  In the actual computation, Lorentzians with the broadening parameter $\epsilon$ are substituted for the $\delta$ functions in eq.~(\ref{eq:sigma-reg}), and the Lanczos diagonalization method is used to obtain $\sigma(\omega)$ and other quantities for finite systems ($N\le 14$) with the open boundary.

\begin{figure}[hbt]
\begin{center}
\includegraphics[width=5.5cm,clip]{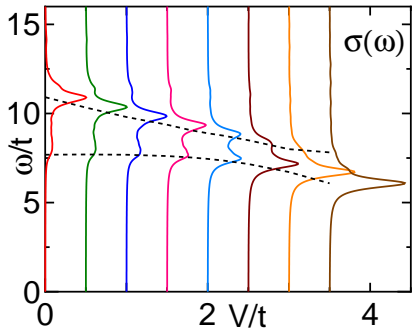}
\end{center}
\caption{ (Color online) Optical conductivity spectra (solid lines) and excitation energies (dotted lines) of photoexcited states for $\delta=0.4,U/t=10$ and $N=12$. The broadening $\epsilon$ is set to $0.3t$.}
\label{fig:sw}
\end{figure}

Figure~\ref{fig:sw} shows $\sigma(\omega)$ and the excitation energies of the two pronounced peaks for $\delta=0.4,U/t=10$, and $N=12$.  Although  $\delta$ is rather small in this case, our argument is confirmed to be appropriate.  For $V=0$, there is a dominant peak at $\omega/t \sim 11$, which corresponds to the intradimer CT exciton state.  As $V$ increases, the excitation energy of this peak decreases linearly, and approaches that of the lowest photoexcited state $|\psi_{1\rm opt}\rangle$.  It has been shown that $|\psi_{1\rm opt}\rangle$ for small $V$ is the interdimer CT state, and its spectral weight is small.  However, as the energy of $|\psi_{1\rm opt}\rangle$ is approached by that of the intradimer CT exciton, the spectral weight increases.  Around $V/t=2$, the gap between the two levels is the smallest, and for larger $V/t$ the lowest state $|\psi_{1\rm opt}\rangle$ has the dominant spectral weight, implying the intradimer CT exciton state.

\subsection{Holon distribution}

\begin{figure}[hbt]
\begin{center}
\includegraphics[width=5.5cm,clip]{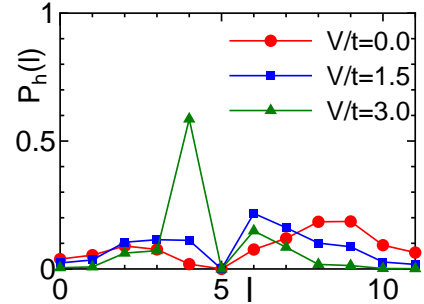}
\end{center}
\caption{ (Color online) Holon distribution function for $\delta=0.4,U/t=10$ and $N=12$.}
\label{fig:Phd04}
\end{figure}

To demonstrate the change of the lowest photoexcited state $|\psi_{1\rm opt}\rangle$ more clearly, we calculate a quantity associated with the nature of the photoexcited state.  Here we consider the holon distribution function $P_{\rm h}$ of $|\psi_{1\rm opt}\rangle$, which is defined by
\begin{equation}
 P_{\rm h}(l) \equiv  \frac{\langle\psi_{1\rm opt}|\hat{P}_{\rm 1hd} (n^{\rm d}_{N/2-1}n^{\rm h}_{l}) \hat{P}_{\rm 1hd}|\psi_{1\rm opt}\rangle}{|\hat{P}_{\rm 1hd}|\psi_{1\rm opt}\rangle|^2 },
\end{equation}
where $\hat{P}_{\rm 1hd}$ is the projection operator onto the subspace that has only one holon-doublon pair, and $n^{\rm d(h)}_l$ is the number operator of doublon (holon) defined by
\begin{equation}
 n^{\rm d}_l \equiv n_{l,\uparrow}n_{l,\downarrow} \quad {\rm and } \quad n^{\rm h}_l \equiv (1-n_{l,\uparrow})(1-n_{l,\downarrow}).
\end{equation}
Here we put the doublon at the $(N/2-1)$-th site, which is the left side of the central bond connecting two dimers.
Figure~\ref{fig:Phd04} shows $P_{\rm h}(l)$ for $\delta=0.4$.   For $V/t=0$, the holon spreads over the whole system, reflecting the unbound nature of the state.  When $V/t$ is introduced, $P_{\rm h}(6)$ increases, and becomes the largest in the whole system at $V/t=1.5$.
 Thus, $|\psi_{1\rm opt}\rangle$ is considered to be the interdimer CT exciton state, where the binding occurs between the holon and the doublon that are generated by the interdimer CT excitation.
 As $V/t$ further increases, the state $|\psi_{1\rm opt}\rangle$ becomes the intradimer CT exciton state where the holon and the doublon lie in a single dimer.

\section{Photoexcited States of TTTA}

Let us apply our theoretical conclusion to understanding the photoexcited states of an organic 
 quasi-1D dimerized Mott insulator, TTTA.~\cite{TTTA2}  The optical properties of TTTA have been studied experimentally, and it has been shown that $\sigma(\omega)$ of TTTA has a double-peak structure.~\cite{fujita} However, the origin of the peak structure has not completely been understood.~\cite{TTTA1,fujita}

\begin{figure}[hbt]
\begin{center}
\includegraphics[width=5.5cm,clip]{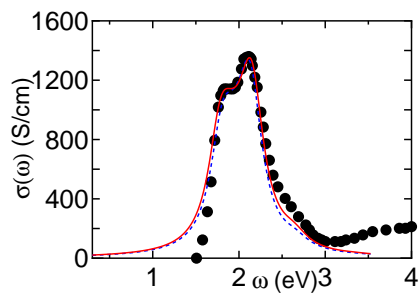}
\end{center}
\caption{(Color online) Comparison between the measured optical conductivity of TTTA from ref.~\citen{fujita} (filled circles) and the calculated optical conductivity of $N=14$ systems with the parameter sets described in the text (solid and dotted lines).}
\label{fig:fujita}
\end{figure}

To address this issue, we first estimate the physical parameters of TTTA.  As for the dimerization $\delta$, $\delta = 0.46$ is obtained on the basis of the extended H\"{u}ckel calculation in ref.~\citen{fujita}.  The other parameters are determined by comparison  between the experimental and the numerical results of $\sigma(\omega)$, and they turn out to be $U/t=10,V/t=1.8$, and $t=0.235$ eV with the Lorentzian width $\epsilon=0.7t$ (see the dotted line in Fig.~\ref{fig:fujita}).  The numerical result well reproduces the double peak around 1.8eV and 2.1eV.  
We also note that the obtained Coulomb interaction $U=2.35{\rm eV}$ is comparable to that by a first-principles calculation ($U=2.9{\rm eV}$),~\cite{ohno} supporting our estimation.
For better fitting of the shoulder structure around 2.5eV, slightly smaller $\delta=0.4$ and $U/t=10,V/t=1.7,t=0.235$ eV with $\epsilon=0.7t$ is also a good parameter set (see the solid line in Fig.~\ref{fig:fujita}). 

\begin{figure}[hbt]
\begin{center}
\includegraphics[width=6.0cm,clip]{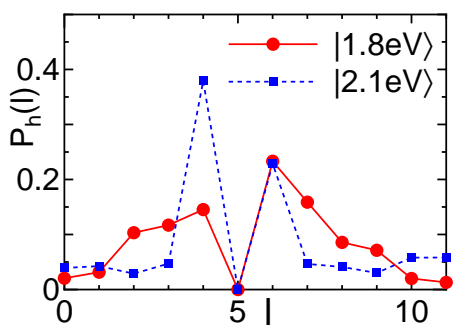}
\end{center}
\caption{(Color online) Holon distribution functions in two photoexcited states $|1.8{\rm eV}\rangle$ and $|2.1{\rm eV}\rangle$ for $\delta=0.4,U/t=10,V/t=1.7$ and $N=12$ with $\epsilon=0.05t$.}
\label{fig:Phd}
\end{figure}

Next, we show that the lowest photoexcited state of TTTA is the interdimer CT exciton state, and the second peak is caused by the intradimer CT exciton state.  To illustrate this conclusion,  we calculate the holon distribution $P_{\rm h}(l)$ of the correction vector~\cite{iwano2}
\begin{equation}
 |\omega\rangle= A \frac{1}{({\cal H} - E_0 -\omega)^2 + \epsilon^2} J|\psi_0\rangle,
\label{eq:psi_w}
\end{equation}
where $A$ is the normalization factor. This formulation facilitates us obtaining photoexcited states with arbitrary $\omega$.  Using eq.~(\ref{eq:psi_w}), we obtain $P_{\rm h}(l)$ of photoexcited states corresponding to the two peaks, $|\omega=1.8{\rm eV}\rangle$  and $|\omega=2.1{\rm eV}\rangle$, and the results for the latter parameter set ($\delta=0.4,U/t=10,V/t=1.7$) are shown in Fig.~\ref{fig:Phd}.  The results clearly show that $|1.8{\rm eV}\rangle$ is an interdimer CT exciton state and  $|2.1{\rm eV}\rangle$ is an intradimer CT exciton state.  This assignment is consistent with the larger spectral weight of $|2.1{\rm eV}\rangle$.

From the experimental point of view, Fujita {\it et al.} have made the same assignment of these peaks; $|1.8{\rm eV}\rangle$ is the interdimer CT exciton state and  $|2.1{\rm eV}\rangle$ the intradimer CT exciton state.~\cite{fujita}. This conclusion is derived from the dominant spectral intensity of $|2.1{\rm eV}\rangle$.  However, the same research group made the opposite assignment; $|1.8{\rm eV}\rangle$ is the intradimer CT exciton state and  $|2.1{\rm eV}\rangle$ the interdimer CT exciton state.~\cite{TTTA1} This was derived from the excitation energies. They have erroneously considered that the intradimer CT exciton would have the larger binding energy because the holon and the doublon are confined in a small region.

Our numerical results support the former conclusion.  In addition, our analysis from the decoupling limit proves that the excitation energy is not always a good measure to distinguish the nature of the photoexcited state in this system.  As shown in eqs.~(\ref{eq:delta_intra}) and (\ref{eq:delta_inter}), the excitation energy of the interdimer CT state can be lower than that of the intradimer CT state for small $V/t$. It is caused by the kinetic energy gain of the interdimer CT state, and this is the case for TTTA.  For large $V/t$, the intradimer CT exciton state becomes the lowest photoexcited state.

\section{Summary}

We have studied the properties of the charge-transfer excitations in the 1D dimerized extended Hubbard model.  Numerical results combined with the analysis from the decoupling limit have proved that there are three classes of photoexcited states: the interdimer CT unbound state, the interdimer CT exciton state, and the intradimer CT exciton state.  Which state is the lowest-energy excitation depends on the nearest-neighbor Coulomb interaction $V$.  This classification of the photoexcited states is applied to the interpretation of the optical conductivity spectrum of the organic material TTTA.  We conclude that the lowest photoexcited state ($|1.8{\rm eV}\rangle$) of TTTA is the interdimer CT exciton state and the higher excited state ($|2.1{\rm eV}\rangle$) is the intradimer CT exciton state.

\section*{Acknowledgments}

The authors are grateful to Prof. H.~Okamoto for enlightening discussions.  This work was supported by Grants-in-Aid for Creative Scientific Research (No.~15GS0216), for Scientific Research on Priority Area ``Molecular Conductors'' (No.~15073224), for Scientific Research (C) (No.~15540354), and Next Generation Super Computing Project, Nanoscience Program, from the Ministry of Education, Culture, Sports, Science and Technology, Japan.  Some of numerical calculations were carried out on Altix3700 BX2 at YITP in Kyoto University, and on TX-7 and PRIMEQUEST at Research Center for Computational Science, Okazaki, Japan.


\end{document}